\documentstyle[preprint,aps]{revtex}
\begin{document}

\draft
\title{Ferromagnetism in the Infinite-$U$
Hubbard Model}
\author{Shoudan Liang and Hanbin Pang}
\address{Department of Physics,\\
Pennsylvania State University,\\
University Park, PA 16802}
\date{\today}
\maketitle
\begin{abstract}

We have studied the stability of the ferromagnetic 
state in the 
infinite-$U$ Hubbard model on a square 
lattice by approximate 
diagonalization of 
finite lattices using the density matrix 
renormalization group
technique.  By studying lattices with up 
to 100 sites, we
have found the 
ferromagnetic state to be stable below the hole 
density of $\delta_{c} = 
0.22$.  Beyond $\delta_{c}$, the total spin of 
the ground state decreased gradually to zero 
with increasing hole density.
\end{abstract}

\newpage
\narrowtext

The origin of many unusual electronic properties of 
high $T_c$ superconductors can be traced to strong  
electron-electron repulsion in the $CuO$ planes.
The Hubbard model is the simplest 
description of such repulsive
interactions. Because of its simplicity 
the Hubbard model plays a role
in many-body problems similar to 
that of the Ising model in 
phase transition problems. However, 
the Hubbard model is
still very difficult to analyze. After forty years of
research, we are still unsure of even its most basic 
features\cite{PWA1,Lieb}.
In this letter, we focus on the infinite-$U$ limit 
to begin looking for unusual behaviors suggested 
recently\cite{PWA,EDWARDS1}. There are several 
reasons for studying this limit.
First of all, the antiferromagnetic state
at half filling in the large $U$ limit is 
incompatible with the  
motion of holes in the metallic phase\cite{kane}. 
It is interesting
to learn about the spin background preferred by 
the motion of the
holes without the complication of the antiferromagnetic  
interaction. A well understood infinite-$U$ limit also
provides a starting point for systematic expansion in $t/U$.
Furthermore, the ground state of the 
infinite-$U$ Hubbard model
at small doping is believed to be ferromagnetic.
This provides a mechanism for itinerant 
ferromagnetism\cite{her}. But it is  
controversial whether there is a finite 
range of hole density where
the ground state is ferromagnetic. 
This letter provides strong evidence that such a finite
region indeed exists for the square 
lattice. We used the recently 
developed density matrix renormalization group(DMRG) method to 
compute the critical hole doping\cite{white}.

The investigation of the infinite-$U$
Hubbard model has a long history.  
The earliest rigorous result is
due to Nagaoka\cite{nag}, and 
independently Thouless\cite{thou0}, 
showing that in the case
of one hole, the ground state on a 
bipartite lattice is the ferromagnetic
state (also known as Nagaoka state), 
where all the spins are aligned in
the same direction.  Since then, 
progress on this difficult problem
has been slow\cite{roth1,roth2,rr,tak,dr,tasaki,trugman}.  
Recently, it has been shown\cite{dw,frds}
that for two holes the Nagaoka 
state is not the ground state.   
However, the proposed two-hole trial 
state\cite{dw} has essentially local
ferromagnetic correlation.  Shastry 
{\em et al.} considered\cite{ska} 
the instability of one spin flip of 
the Nagaoka state at finite hole 
density. They shown that the Nagaoka 
state is unstable when the  
hole density exceeds $\delta_c = 0.49$.  
This result has been 
improved\cite{basilelser} to yield $\delta_c = 0.41$. 
The single spin flip state has also been studied by
von der Linden and Edwards\cite{wvdl} using a more
general trial wave function. They shown 
that the Nagaoka state is
unstable against a single spin flip for $\delta > 0.29$.

By comparing the high temperature
expansion coefficients of the infinite-$U$ 
Hubbard model with that of  
a free spinless fermion Hamiltonian, 
Yedidia\cite{yedidia} conjectured that the transition to
the non-ferromagnetic state occurs at $\delta \simeq 3/11$.  
The high temperature expansion has 
been extended to higher order by
Putikka {\it et al.}\cite{putikka}. 
When the free energy is  
extrapolated to zero temperature, 
their calculation suggests $\delta_c =0$.

An extensive exact diagonalization 
investigation\cite{ry2,bry} has been
carried out using Lanczos method, which is limited
to small clusters. A very large
finite-size effect has been observed. 
On a square lattice with periodic
boundary conditions, the Nagaoka state is stabilized for
the close-shell configurations when the number of holes
is 1, 5, 9, .... At other hole fillings
the Nagaoka state tends to be 
destabilized on small lattices
because the energy change from one 
shell to the next is too large.
Because of this, the ground-state 
magnetization oscillates with
the number of holes\cite{ry2}.

We have studied the stability of the Nagaoka state as
well as the nature of the transition to the paramagnetic state
by approximate diagonalization
on finite lattices. The recently developed density matrix
renormalization group method by White\cite{white} and our 
own extension\cite{liang} to
two dimensions allow us to perform
calculations on much larger lattices than previously
possible and with high accuracy.

For $L_x\times L_y$
lattices, when $L_{y}>2L_{x}$ the 
difficulty of the calculation
depends only weakly on $L_{y}$\cite{liang}.  
By making $L_{y}$ suitably long,
spacing between the nearest $k_y$ can be 
made as small as we want and thus 
the close-shell effect can be eliminated.

Based on diagonalizations of 
$L_x\times L_y$ lattices with $L_y=20$,
we find the critical hole concentrations for the
onset of instability in the Nagaoka state
to be almost the same for $L_x=2,3$ and $4$. This  
suggests that the critical hole doping 
we calculated, $\delta_c=0.22$, is  
close to the
bulk limit.  This value is close to 
but lower than $\delta_c=0.29$ obtained from Edwards trial
wave function\cite{wvdl} for the 
case of single spin flip.  We show 
in contrary to previous finding\cite{putikka} 
that a finite region of hole
doping exists below $0.22$ where 
the fully ferromagnetic state is stable.

We calculate the ground-state energy of small clusters 
using the DMRG method, in which one reduces
degrees of freedom by keeping the eigenstates
of the density matrix\cite{white}. 
This is in contrast to the conventional
real space renormalization group method where 
the low energy eigenstates of the block Hamiltonian
are kept. An iterative 
procedure\cite{white} is used to systematically
improve the approximation to the density matrix. 
The DMRG method proves to be highly accurate
for one dimensional systems. For quantum spin chains, the 
ground-state energy
can be calculated to a high 
accuracy of $10^{-6}$\cite{white}.
When the method is applied to 
the quasi-one dimensional system of several
coupled chains\cite{liang}, 
the number of states needed to compute
the energy to a fixed accuracy 
grows exponentially with the
number of chains, but is independent 
of the length of the chains.
It can also be shown\cite{liang} that the energy calculated
in the finite cluster DMRG method always provides
a variational upper bound to the ground-state energy.

We study the one-band Hubbard model with $U=\infty$ on
$L_x\times L_y$ square lattices with free boundary
conditions in both directions.  We are restricted to small 
$L_x$ because the accuracy of the DMRG method deteriorates 
at large $L_x$. In this work, the calculations are done on 
strips with $L_x=2,3,4, 5$ and $L_y=20$. The large value of 
$L_y$ used reduces finite 
size effects due to the {\bf k}-space shell closing discussed 
previously.  

Let $E_N(Q,S_z)$ be the energy calculated for the system with
$Q$ holes (the number of electrons 
is $N-Q$) and total $z$-direction
spin $S_z$ on an $L_x\times L_y$ lattice with $N$ sites.
The critical hole doping $\delta_c$ is determined 
by comparing $E_N(Q,S_z=0)$ with the
energy of the Nagaoka state, $E_{nag}(Q)$, 
which is the energy of
$N-Q$ spin up electrons on the same lattice. (We assume
the number of electrons $N-Q$ is even.
When $N-Q$ is odd, set $S_z=1/2$.)
Because of the global SU(2) symmetry of the Hubbard
model, the Nagaoka state with 
total spin $S=(N-Q)/2$ is $(2S+1)$-fold
degenerate. One of these states has $S_z=0$. Since
the DMRG method calculates a variational upper bound
to the ground-state energy, we 
have $E_N(Q,S_z=0)\ge E_{nag}(Q)$,
{\it if} the ground state is the Nagaoka state.
On the other hand, if $E_N(Q,S_z=0) < E_{nag}(Q)$
the ground state is not the Nagaoka state.
The smallest hole doping for which this occurs
determines the critical doping $\delta=Q/N$.
Since energy computed in the 
DMRG method is a variational upper  
bound, the critical doping
$\delta_c$ we estimated from 
the condition $E_N(Q,S_z=0) < E_{nag}(Q)$
is an upper bound to the true $\delta_c$.

Since below $\delta_c$ the exact $E_N(Q,S_z=0)$ is equal 
to $E_{nag}(Q)$, the difference between
the actual $E_N(Q,S_z=0)$ calculated and the 
corresponding Nagaoka energy provides
an estimate for the accuracy of our calculations.
The accuracy of our calculations 
in the relevant doping region varies from
0.03\% for $2\times20$(with $M=52$) to 0.5\% for $4\times20$ 
lattices(with $M=102$)(Fig.\ref{chains}), where
$M$ is the number of states kept.

In Fig.\ref{chains}, the energy 
difference $E_{N} (Q,S_{z}=0) - E_{nag}(Q)$ 
between the calculated energy and 
the Nagaoka energy is shown as a 
function of hole doping ($S_{z}=1/2$ 
if the number of electrons is odd).
The energies are normalized 
to $E_{nag}(Q)$. At critical doping $\delta _{c}$,
the energy difference 
changes from positive to negative. 
The calculated energy 
$E_{N} (Q,S_{z}=0)$ reported here 
are for the largest number of 
internal states kept.  We have not attempted the
extrapolation to the infinite-$M$ limit 
because we are uncertain about the 
validity of such an extrapolation and 
because at the largest $M$, 
$E_{N}(Q,S_{z})$ gives a nice 
variational upper bound. For $\delta < 
\delta _{c}$, the energy difference 
is positive and flat.  For $\delta >
\delta _{c}$ the energy difference 
turns negative abruptly and decreases
linearly with $\delta - \delta _{c}$(at 
least for $L_{x} = 2, 4$).  In 
Fig. 1(a), the data for $2\times20$ and $2\times30$ 
are almost indistinguishable from each other
indicating that $L_{y} = 20$ is large enough.
Since the number of states needed for 
calculations with fixed accuracy 
grows exponentially with the $L_{x}$, the errors 
for $5\times20$ (Fig.\ref{chains}(d))
are considerably larger.  Because of 
this, the energy difference becomes 
negative at a higher doping.

The condition $E_N(Q,S_z=0)\ge E_{nag}(Q)$
is a necessary condition for the 
stability of the Nagaoka state.
It only suggests but does not prove the ground state
is ferromagnetic. However,
for the $2\times20$ and $2\times30$ lattices 
when hole doping is smaller than 0.22, 
$E_N(Q,S_z=0)- E_{nag}(Q)$ is as 
small as $10^{-5} E_{nag}(Q)$
which strongly suggest that the 
true $E_N(Q,S_z=0)$ is in fact
equal to $ E_{nag}(Q)$. The ferromagnetic state is at
least a degenerated ground state.
The similarity between the data for $L_{x}=3,4$ 
and $L_{x}=2$ suggests that the Nagaoka state
is stable below about 20 percent doping.
Also, the critical hole dopings 
change very little for $L_x=2,3,4$. 
This insensitivity indicates that $\delta_c=0.22$ is
close to the bulk limit.

Near $\delta_c$ we also calculated 
the energy of the lowest state
with one spin flipped ($S_z=\frac{(N-Q)}{2}-1$). We expect to
achieve higher energy accuracy because 
the dimensions of the Hilbert
space is reduced from the $S_z=0$ case. 
We have verified that for $L_x=2,3,4$,
the $\delta_c$ inferred from the energy 
with one spin flipped
is the same as the $S_z=0$ case.

To investigate the effects of lattice shape anisotropy on
the critical doping $\delta_c$, we 
introduced hopping anisotropy:  
$t_x=0.5$ and $t_y=1$ on $L_x\times L_y$ 
lattice with $L_x=4$  
$L_y=20$. Remarkably,
the critical doping for this 
system(Fig.\ref{tx}) is very close to $\delta_c =  
0.22$ of the isotropic case 
(when $t_x=t_y=1$ in Fig.1(c)). This insensitivity
to hopping anisotropy gives us 
some confidence that $\delta_c=0.22$
is close to the bulk limit.

We now discuss the nature of the 
ferromagnetic to paramagnetic 
transition after the doping exceeds 
$\delta _{c}$. There are two 
possibilities: ({\it i}) the
total spin $S$ of the ground state 
changes discontinuously from the maximum  
$\frac{N-Q}{2}$ to zero, or
({\it ii}) as the hole concentration 
$\delta$ exceeds the critical
doping, the ground-state total spin 
reduces gradually to zero as
$\delta$ is increased. We can in 
principle distinguish between 
these two possibilities by computing the ground state
energy $E_{N}(Q, S_{z})$ as a 
function of $S_{z}$.  In case ({\it i}), the energy
$E_{N}(Q, S_{z})$ decreases until $S_{z}$ reaches zero.  For
case ({\it ii}), $S_{z}$ stops decreasing at $S_{c}(Q)$ and
$S_{c}(Q)$ goes to zero when $Q$ is increased.

Our data is consistent with case ({\it ii}) above namely 
that there exists a doping region 
$\delta _{c} < \delta < \delta_{c_{1}}$
where the ground-state total spin is 
between $S_{max}=\frac{N-Q}{2}$ and 
zero.  For $\delta > \delta _{c _{1}}$ 
the ground-state total spin 
becomes zero.  Fig.\ref{sz} 
shows some representative data. In 
Fig.\ref{sz}(a) for $\delta = 0.3$ on 
$2\times30$ lattice, the 
ground-state energy $E(Q, S_{z})$ drops 
quickly with decreasing $S_{z}$ 
until $S_{z} = 0.5S_{max}$.  After that the 
energy is flat.  The total spin of the 
ground state is then $S=0.5S_{max}$. The 
slight increase in the energy from $S_{z} = 0.5S_{max}$ 
to $S_{z} = 0$  is due  
to the increased Hilbert space at small $S_{z}$ which makes the 
calculation less accurate.  At $\delta = 0.5$ 
( Fig.\ref{sz}(b)), the energy 
decreases continuously to $S_{z} = 0$.  
This implies that the ground state has zero total spin.  
Similar behaviors are observed for $L_x=4$.
We are unable to determine $\delta_{c_1}$ accurately.
But it is close to 0.40.

We now discuss technical details of the DMRG calculation
specific to the infinite-$U$ limit.  
A general discussion of DMRG procedures for
quasi-one-dimensional systems can be found in 
Ref.\cite{liang}.
The chief computational advantage of the infinite-$U$ limit
over the full Hubbard model is the reduced
Hilbert dimensions.
When expanding a block, we add three states 
per site (empty, spin up and spin down).

The one dimensional system is used 
to initialize the environment blocks
\cite{white,liang}.
One particular difficulty in the infinite-$U$ limit 
is that in one dimension 
all the spin configurations have exactly the same energy.  
The total angular momentum is therefore
undefined.  In the quasi-one-dimensional 
system, however, this degeneracy is
lifted and the ground state has well 
defined total angular momentum.  
One can get around this problem by 
starting from the one dimensional 
Hubbard model with large $U$ which lifts the degeneracy.

Typically six iterations are performed 
for each of several values of $M$ 
(number of internal states kept) 
starting from small $M$.  To preserve 
the $SU(2)$ symmetry, we always keep
the states with the same weight so 
that the actual number of states kept
may be larger than the assigned $M$.

The programming for the DMRG method is much
more complex than Lanczos exact diagonalization.  Our
code for the two dimensional 
Hubbard model contains over 4000 lines.
A crucial issue is how to make sure that the
computer program is correct. Our
computer code has passed two non-trivial tests.
({\it i}) Hubbard model satisfies a 
global spin $SU(2)$ symmetry.  
For states with zero total spin,
the internal states retained should exhibit the
$SU(2)$ symmetry.  In particular, 
the states come in $2S+1$ multiplets,
{\it i.e.} whenever we have a state 
with $z$-component of
spin $S_{z}=S$ we should also find 
states having identical weight (the
diagonals of density matrix)
with $z$-direction of spin being $S-1, S-2, ..., -S+1, -S$.
({\it ii}) For spin polarized state with $S_z={{N-Q}\over2}$, 
the computed energy approaches the exact
answer.

In conclusion, we have studied the 
stability of the Nagaoka state in the
infinite-$U$ Hubbard model in two dimensions
using the density matrix renormalization 
group method. We found
the ferromagnetic state to be stable 
for a finite doping range
near half filling. By computing energy upper bounds
on $L_x\times20$ lattices with $L_x$ up to $5$,
we have shown that the Nagaoka
state becomes unstable for hole doping larger than
22 percent. The ground-state 
total spin decreases gradually
as the hole doping is increased and becomes zero for
more than about 40 percent doping.

The work was supported in part by 
the Office of Naval Research
Grant No. N00014-92-J-1340.

\begin{figure}
\caption{The critical doping $\delta_c$ is obtained by 
comparing a variational upper bound of energy, E,
calculated at doping $\delta$ and $S_z=0$ 
(and $S_z=1/2$ if the number of electrons is odd)
with the energy of Nagaoka state, $E_{nag}$.
The energy difference, normalized to the Nagaoka
energy, is plotted as a function of hole doping.
The Nagaoka state becomes unstable when the energy difference
is negative.
(a) $2\times20$ and $2\times30$ 
lattices. The number of states
kept in the calculation,$M= 52$. The energy accuracy
of the variational bounds are $0.03$ percent. (b)
$3\times20$ lattice with $M$=62. 
(c) $4\times20$ lattice with $M$=102.
(d) $5\times20$ lattice with $M$=120.
}
\label{chains}
\end{figure}

\begin{figure}
\caption{Same as Fig.1(c) but with 
anisotropic hoping $t_x=0.5$
(in the short direction of the lattice) 
and $t_y=1$. $M$ is 120.}
\label{tx}
\end{figure}

\begin{figure}
\caption{The ground-state energy $E(S_z)$ as a function
of $S_z$ calculated at $M$=62. (a) At doping $\delta=0.3$,
the energy decreases as the spins are flipped from 
the Nagaoka state until $S_z\simeq0.5S_{max}$, where
the $S_{max}$ is the spin of the Nagaoka state. The total
spin of the ground state is then close to $0.5S_{max}$.
(b) At larger doping $\delta=0.5$, the energy
decreases continuously. The ground-state total spin is
zero.}
\label{sz}
\end{figure}

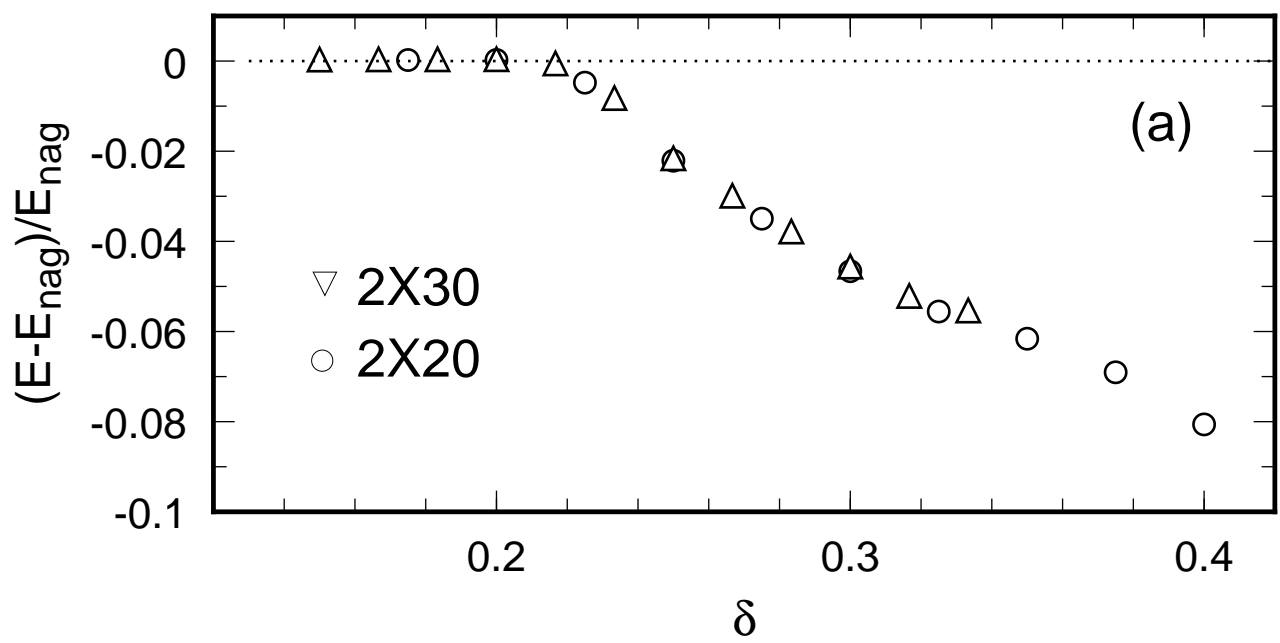
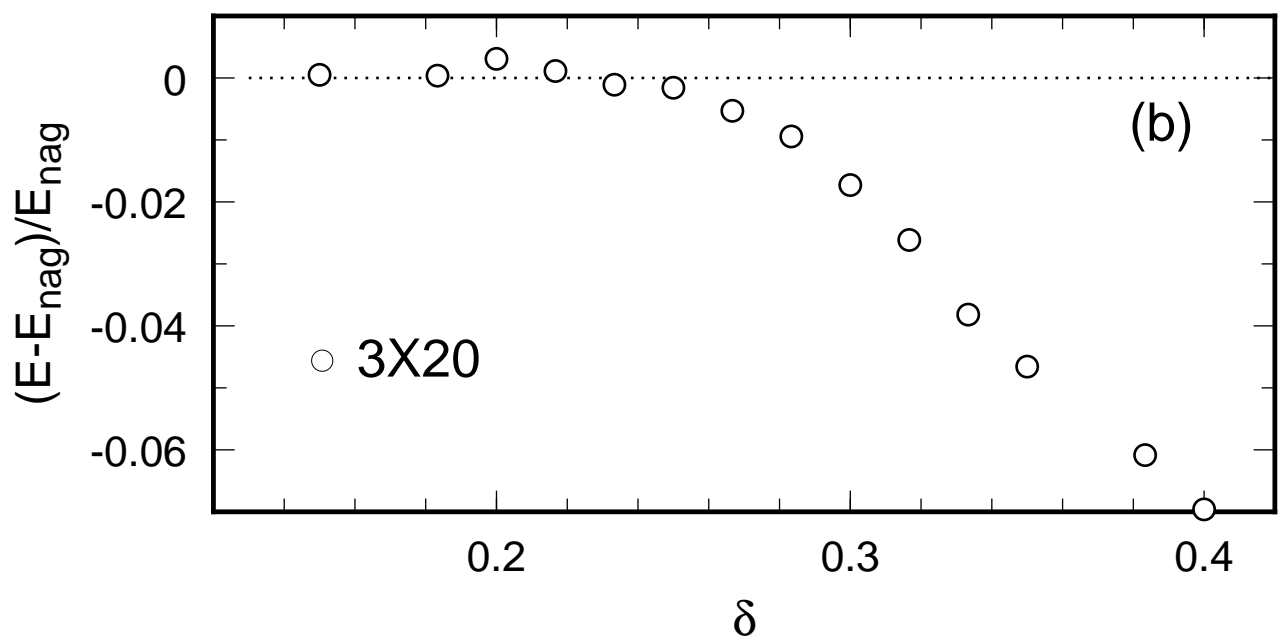
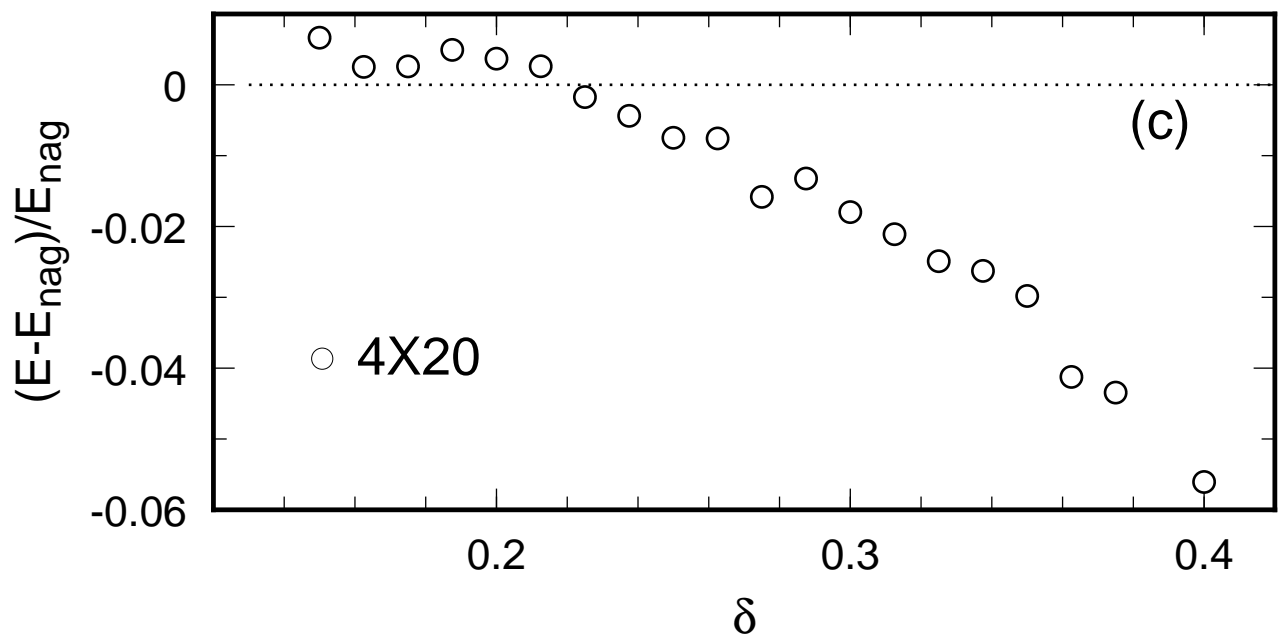
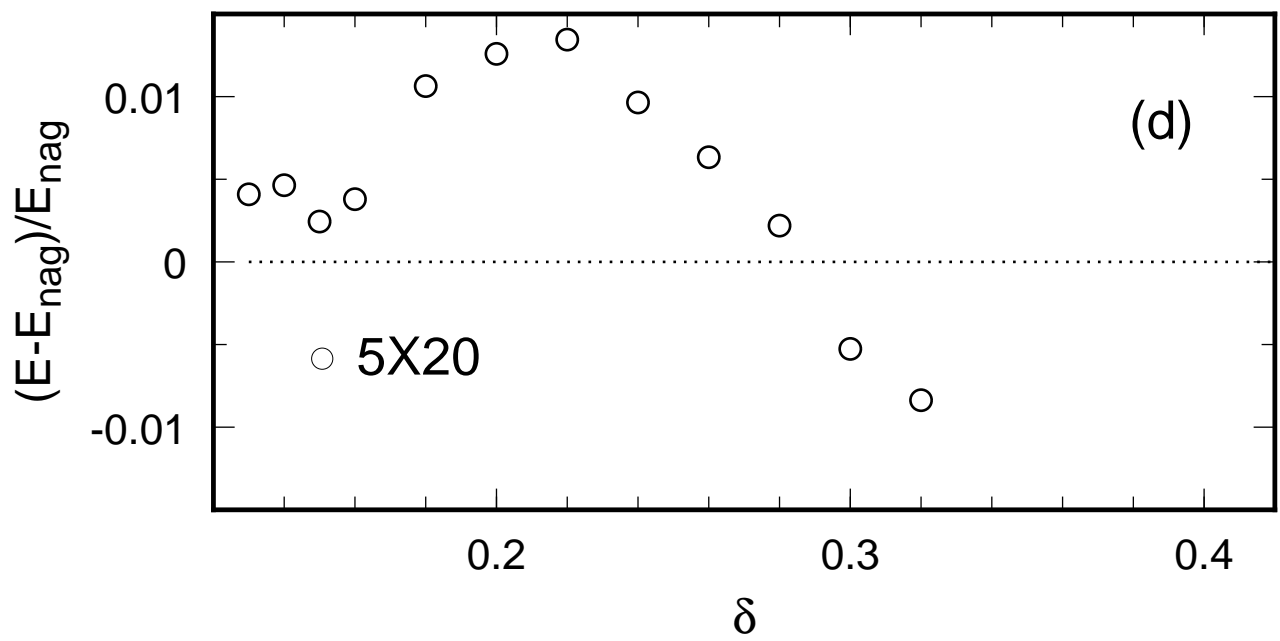
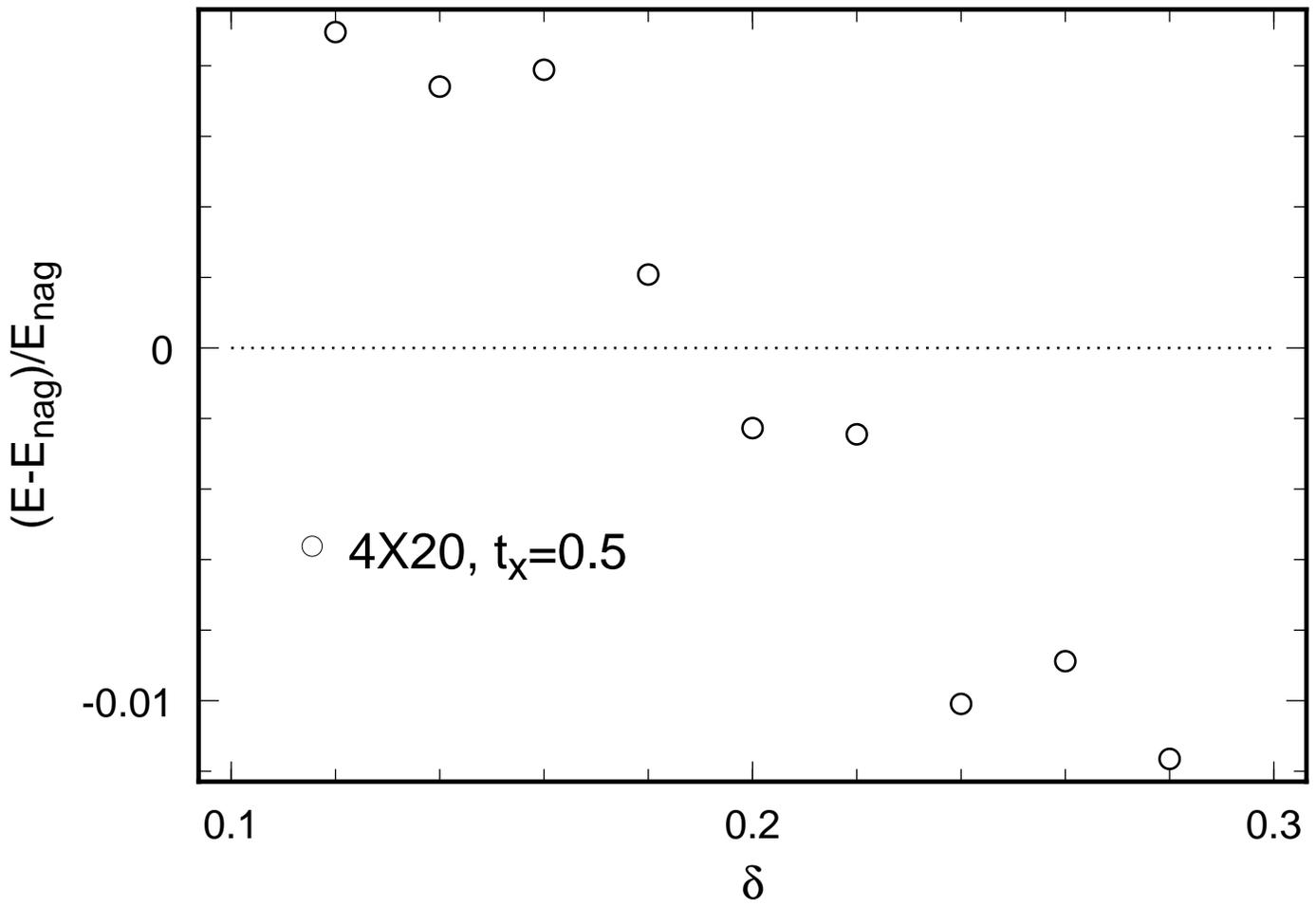
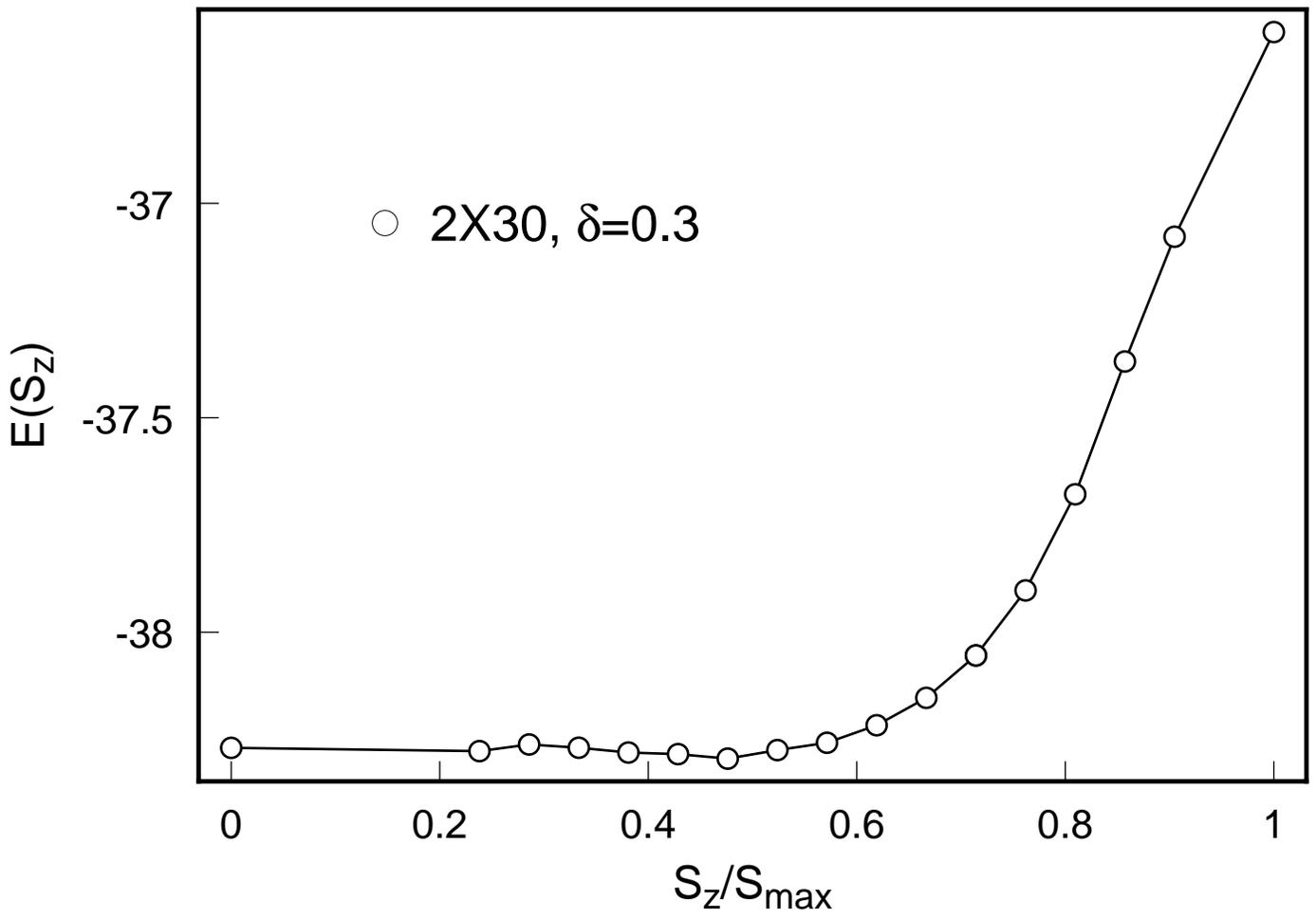
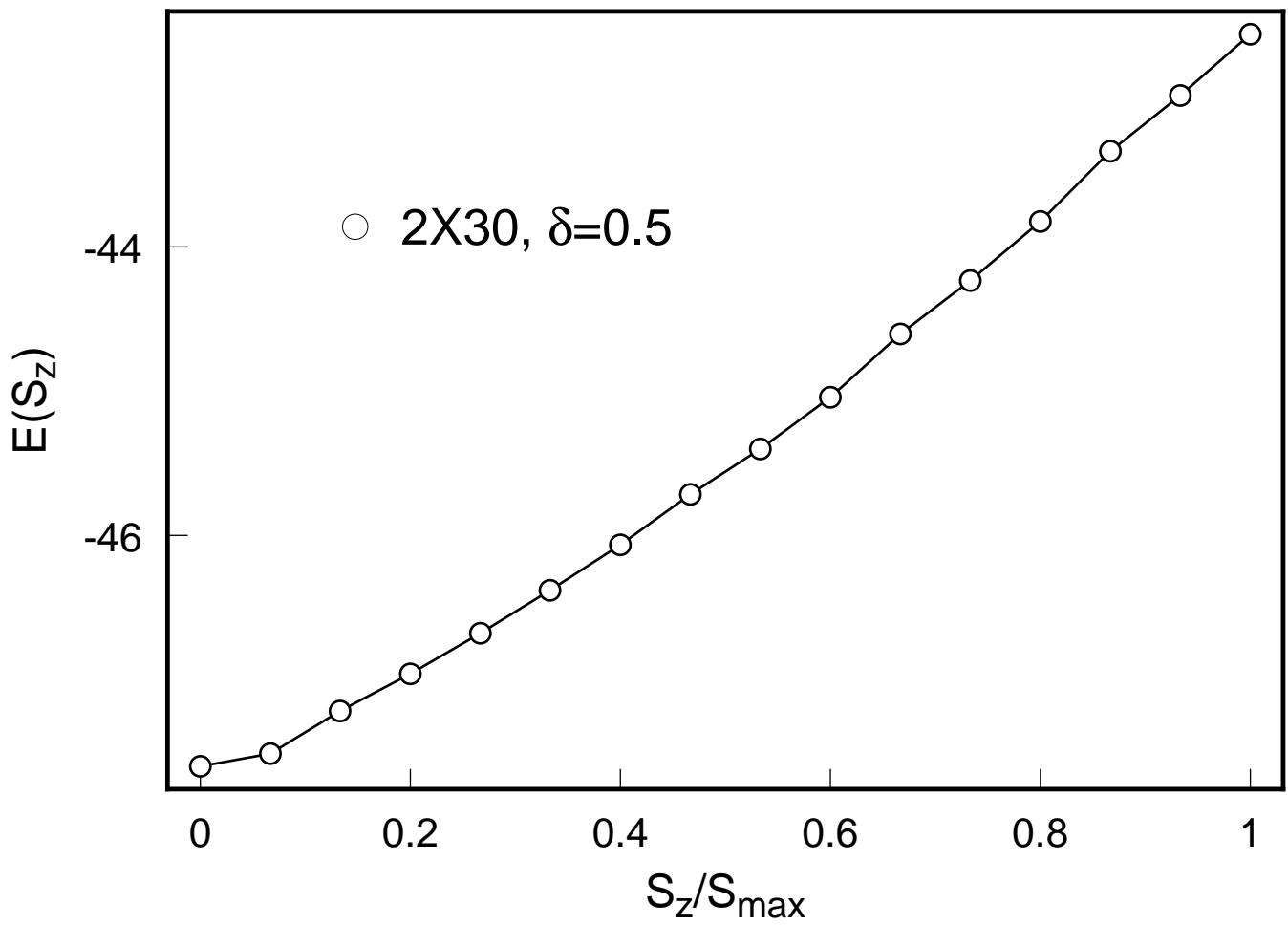
\end{document}